\documentclass[aps,preprint]{revtex4}

\usepackage{graphicx}

\begin{document}

\def \ba {\begin{eqnarray}}
\def \ea {\end{eqnarray}}
\def \be {\begin{equation}}
\def \ee {\end{equation}}
\def \tight {\!\!\!}
\def \ra {\rightarrow}
\def \bscco {Bi$_2$Sr$_2$CaCu$_2$O$_{8 +\delta}$}
\def \ybco {YB$_2$Cu$_3$O$_{7-x}$}
\def \ttju {$t$-$t'$-$J$-$U$ }
\def \tju {$t$-$U$-$J$ }
\def \ttuv {$t$-$t'$-$U$-$V$ }
\def \vr {\vec r}
\def \vi {\vec i}
\def \vk {\vec k}
\def \wt {\widetilde}
\def \wh {\widehat}

\title{Zn impurity induced moments and tunneling conductance
asymmetry in cuprate superconductors}
\author{Jaesung Yoon, Tae-Hyoung Gimm, Hyun C. Lee, and Han-Yong Choi}
\affiliation{Department of Physics, BK21 Physics Research Division,
and Institute for Basic Science Research, \\
Sung Kyun Kwan University, Suwon 440-746, Korea.}

\date{\today}

\begin{abstract}

    The effects of a non-magnetic Zn impurity substituting an in-plane
    Cu are studied by solving the Bogoliubov-de Gennes equation
    self-consistently which is derived from the \ttju Hamiltonian
    with all the allowed order parameters included. The Zn impurity,
    modeled in terms of a potential scatterer in unitary limit,
    induces local staggered magnetic moments around itself, and the
    calculated NMR shifts from the induced moments are in agreement
    with the experimental Cu NMR spectra. We also note that the
    experimentally observed negative slope of the tunneling
    conductance can result from the next-nearest hopping $t'$.

\end{abstract}


\maketitle

After more than a decade of intensive research on the high $T_c$
superconductors, we are still far from a coherent understanding
of the underlying physics \cite{ginsberg}. One informative route
of study is provided by the impurities like Zn and Ni, which
substitute for Cu sites in the CuO$_2$ planes where the
superconductivity (SC) is believed to emerge from. Therefore, the
impurity induced information about the SC and magnetism and their
correlation can provide a clue to the underlying physics of the
high $T_c$ SC in that a common feature of the cuprates is the
proximity of the antiferromagnetism and SC. The high resolution
atomic scale probes on the impurity substituted cuprates provide
important information in this regard. For instance, Pan $et~al.$
reported the scanning tunneling microscopy (STM) study on a Zn
substituted \bscco, that is, the tunneling conductance as a
function of the atomic position and bias voltage, $G( \vr, V)$
\cite{pan}. They observed among others that (a) at $\vr =$ Zn
site, there is a strong peak in $G(\vr,V)$ as a function of $V$
near zero bias voltage ($V \approx -1.5$ meV), (b) at $V=-1.5$
meV, the real space distribution of the intensity $G(\vr, V)$
shows local maxima at next-nearest Cu sites from the Zn, and (c)
the spatially averaged conductance $\int d \vr G(\vr , V)$ has a
negative slope as a function of $V$. Also, several groups
reported NMR experiments on Zn \cite{alloul,julien}, Li
\cite{bobroff}, or Ni \cite{williams} substituted \ybco. It was
found that (d) local staggered moments were induced and SC was
destroyed around the impurity site, and away from there the
induced moments become vanishing and the pairing amplitudes heal
back to the bulk value.

Interpretation of these observations at present is controversial.
For the observations (a) and (b), some groups proposed the
blocking or interference effects \cite{balatsky,ting2}, but other
groups introduced the Kondo effect due to the impurity induced
magnetic moments \cite{sachdev}. For the observation (c), Rantner
and Wen suggested from a $SU(2)$ slave boson mean-field
calculation that the asymmetric tunneling background may be due
to the strong onsite repulsion \cite{rantner}. Although there are
quite a few theoretical works on the magnetic or non-magnetic
impurity effects on superconductivity
\cite{balatsky2,salkola,salkola2,chchoi}, no theoretical
calculation of the induced magnetic moments due to a non-magnetic
impurity in a SC state has been reported yet. We report such a
calculation in this paper. We study the \ttju Hamiltonian with
the Zn impurity modeled as a potential scatterer near the unitary
limit to understand the observations (c) and (d) listed above.

In order to address the observations (c) and (d), we
consider the \ttju model given as follows:
 \ba
H &=&  \sum_{\ll i, j \gg \sigma} \left[ (-t_{ij} -\mu\delta_{ij})
c_{i \sigma}^+ c_{j \sigma}+h.c. \right] +  J \sum_{<i, j>}
\left( \wh{\vec{S_i}} \cdot \wh{\vec{S_j}} - \frac{\wh{n}_i
\wh{n}_j} {4} \right) \nonumber \\
&+& U \sum_i \wh{n}_{i\uparrow} \wh{n}_{i\downarrow} + V_{imp}
\sum_{\sigma} c_{i_0 \sigma}^{+} c_{i_0 \sigma}, \label{tju}
 \ea
where $\wh{n}_{i\sigma} = c_{i\sigma}^+ c_{i\sigma}$, $\wh{n}_i =
\sum_{\sigma} \wh{n}_{i\sigma} $, $\wh{\vec{S_i}}$ the spin-1/2
operator, $\mu$ the chemical potential, $U$ the onsite Hubbard
repulsion, $V_{imp}$ the impurity potential at the Zn impurity
site $i_0$, and $J$ is the nearest neighbor exchange integral.
The $<>$ and $\ll \gg$ stand for the summations up to,
respectively, the nearest and next-nearest neighbors, and $t_{ij}
$ is equal to $t$ for the nearest and to $t'$ for the
next-nearest neighbors. The normal state band structure is then
given by
 \ba \epsilon_k = -2t  [ \cos(k_x a) +\cos(k_y a) ] - 4t'
\cos(k_x a) \cos(k_y a) . \label{band}
 \ea
Note that we included both $U$ and $J$ in the Hamiltonian. This
was motivated by the recent work by Daul {\it et al.} who
suggested that the \tju model be more appropriate than the
$t$-$J$ or Hubbard models to describe the real cuprate system by
allowing the charge fluctuations with the finite Hubbard $U$
\cite{daul}.

The \ttju model of Eq.\ (\ref{tju}) was treated with the
Hartree-Fock-Bogoliubov (HFB) approximation to yield the
Bogoliubov-de Gennes (BdG) equation in the real space. It was
then solved via iterations with all the allowed order parameters
determined self-consistently, that is, the $d$-wave pairing,
bond, charge, and spin order parameters. We found that (1) local
antiferromagnetic moments are induced due to a non-magnetic
impurity, (2) the tunneling conductance shows a negative slope as
a function of bias voltage $V$ due to the next-nearest neighbor
hopping $t'$, and (3) the local density of states (LDOS) at the 4
nearest neighbors shows a peak near zero energy. The calculated
NMR spectra from the distribution of the induced magnetic moments
is found to be in agreement with the experimentally observed
$^{63}$Cu NMR spectra \cite{julien}. There is some controversy in
understanding the experimental observations (a) and (b) with the
result (3) as mentioned before. Here, we will report the results
(1) and (2) in more detail to understand the observations (c) and
(d).

The BdG equation derived from the Hamiltonian of Eq.\ (\ref{tju})
within the HFB approximation can be written as
 \ba
\sum_j \left( \begin{array} {cc} H_{ij\downarrow} & \Delta_{ij}\delta_{i+\tau,j} \\
\Delta_{ij}^* \delta_{i+\tau,j} & - H_{ij\uparrow} \end{array}
\right) \left(\begin{array}{c} u_{j,n}\\ v_{j,n}
\end{array} \right)
= E_n \left( \begin{array}{c}{u_{i,n}}\\{v_{i,n}}
\end{array} \right),
\label{bdg}
 \ea
where
 \ba
H_{ij\sigma} &=& -(t +J\chi_{ij, -\sigma}^*)\delta_{i+\tau,j} + (U
n_{i,-\sigma} -\mu) \delta_{ij}
\nonumber \\
&-& J n_{i, -\sigma} \delta_{i+\tau,j} - t'\delta_{i+\tau',j}
+V_{imp} \delta_{i,i_0},
 \ea
and the subscripts $\tau$ and $\tau^{\prime}$ are,
respectively, the nearest and next-nearest neighbors.
The order parameters, $d$-wave pairing $\Delta_{ij}$, bond-order
$\chi_{ij \sigma}$, spin up and down charge densities
$n_{i\sigma}$, were determined from the
following self-consistency conditions:
 \ba
\Delta_{ij} &=& J(\left< c_{i\uparrow} c_{j\downarrow} \right> +
\left< c_{j\uparrow} c_{i\downarrow} \right> )
= J \sum_n [u_{j,n} v_{i,n}f(-E_n) + u_{i,n}v_{j,n}f(-E_n)], \nonumber\\
\chi_{ij\uparrow} &=& \left< c_{i\uparrow}^+ c_{j\uparrow} \right>
= \sum_n u_{i,n}^* u_{j,n} f(E_n), \nonumber \\
\chi_{ij\downarrow} &=& \left< c_{i\downarrow}^+ c_{j\downarrow} \right>
= \sum_n v_{i,n}^* v_{j,n} f(-E_n), \nonumber \\
n_{i \uparrow} &=& \left< \wh{n}_{i \uparrow}\right>
= \sum_n u_{i,n}^* u_{i,n}f(E_n), \nonumber \\
n_{i \downarrow} &=& \left< \wh{n}_{i \downarrow}\right> =\sum_n
v_{i,n}^* v_{i,n}f(-E_n), \label{order}
 \ea
where $\left< \cdots \right>$ represents the thermodynamic
average and $f(\omega) = 1/[1+\exp(\beta \omega)]$ is the Fermi
distribution function. Charge and spin densities are then given
by $n_i = n_{i\uparrow} +n_{i\downarrow}$ and $S_z (\vec i ) =
\frac{1}{2} (n_{i\uparrow} -n_{i\downarrow} )$, respectively.
Eq.\ (\ref{bdg}) with the self-consistency conditions of Eq.\
(\ref{order}) was solved via iterations with a unit cell of $20
\times 20$, or $30 \times 30$ sites using a periodic boundary
condition with $5 \times 5$ wave-vector points in the first
quadrant of the Brilouin zone. The LDOS which is proportional to
the tunneling conductance, and local magnetic moments of the
$i$-th site can be obtained from
 \ba
D({\vi},\omega) \tight &=&\tight  \sum_{n, k } \left[ | u_{i,nk}
|^2 \delta(\omega-E_{nk} ) + |v_{i,nk} |^2 \delta(\omega +E_{nk}
)\right],
\nonumber \\
S_z ({\vi} ) \tight &=& \tight \frac{1}{2} \sum_{n,k }
\left[ |u_{i,nk} |^2 f(E_{nk} ) - | v_{i,nk} |^2 f(-E_{nk} )
\right].
 \ea

\begin{figure}
\includegraphics*[scale=0.35]{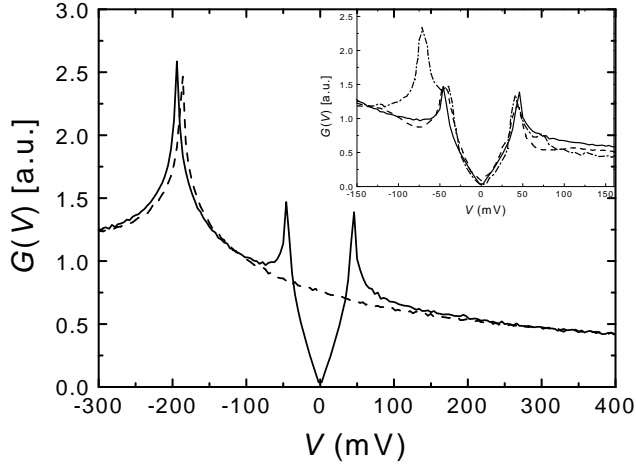}
\caption{ The tunneling conductances in normal (dashed) and SC
(solid curve) states. In the inset, TDOS in SC state in the main
figure is shown in more detail around the gap feature and is
compared with the Pan {\it et al.}\ experiment (dashed curve) and
Rantner and Wen calculation (dot-dashed). }
\end{figure}

A few remarks about the Hamiltonian are in order. The bond-order,
although not essential, was included because (a) it is an allowed
order parameter in the \ttju model unlike in a model with the
onsite repulsion $U$ and next-nearest-neighbor attraction $V$
(\ttuv, or minimal model), and (b) it was found important for
stablizing the bond-centered stripes where the more sophisticated
methods like the density matrix renormalization group predict so
and the half-filled metallic stripes as observed experimentally
\cite{gimm}. We wished to use a model of more general validity
applicable for both the impurity and stripe problems. This is the
main reason we employed the \ttju model in the present study. On
the other hand, the physics discussed here may also be described
in terms of the minimal model because the bond-order is not
essential for the moment induction.

Let us first consider the asymmetric tunneling background of the
observation (c). We take the parameters $t=0.4$ eV, $t' = -0.45
~t$, $U=3~t$, and $J=0.75~t$, which are appropriate for \bscco.
Rather a large value of $J$ was necessary to fit the pairing
amplitude $\Delta = 48$ meV observed in the Pan $et~al.$ STM
experiment. To illustrate the point in a simple way, we put $S_z
=0 $. Fig.\ 1 shows the tunneling density of states (TDOS) in
normal (dashed) and SC state (solid line) at the 16 \% doping
concentration. The two-dimensional tight-binding band given by
Eq.\ (\ref{band}) has a van Hove singularity (vHS) due to a
logarithmic divergence. The vHS, or a peak in DOS occurs at the
band center for $t'=0$. But, for $t' <0$ the peak in the DOS
moves to a lower energy relative to the Fermi level and,
consequently, the DOS has a negative slope around the Fermi level
($\omega=0$) as a function of the bias voltage as shown in the
Fig.\ 1. Small offset of the vHS peaks between the normal and SC
states are due to the pairing and the difference in the bond
orders, and the V-shape around $\omega=0$ of the solid curve is
due to the $d$-wave pairing. This is shown in more detail around
the gap feature in the inset. The solid, dashed, and dot-dashed
curves represent, respectively, the present calculation, Pan
$et~al.$ data, and Rantner and Wen calculation. The agreement of
the present calculation with the Pan $et ~al.$ data is quite
satisfactory without fine adjustments. It seems natural,
therefore, to understand the observed negative slope of the
tunneling background \cite{pan,yaz} in terms of the particle-hole
symmetry breaking around the band center because of the
next-nearest hopping integral $t'$. Rantner and Wen, on the other
hand, argued that the asymmetric tunneling background can be
explained in terms of $SU(2)$ slave-boson approach
\cite{rantner}. But the present explanation is more
straightforward, and fits the experimental observation much
better than their calculation. Interactions may bring in some
corrections in the tunneling conductance but the background slope
will be mainly set by the particle-hole symmetry breaking due to
the $t'$. We therefore believe that the asymmetric tunneling
background is most probably due to the next-nearest hopping
integral $t'$.

Now we turn to the main result of the present work, that is, the
magnetic moment formation due to a Zn impurity substituting an
in-plane Cu site of the cuprate superconductors. We took $J=t$,
$T=0.0005~t$, and all other parameter values the same as in Fig.\
1. A slightly larger value of $J$ was taken to show the doping
induced features more clearly. This set of parameters yields the
pairing $\Delta$ = $0.312 ~ t $, magnetic moment $S_z (\vi )= 0$,
and bond order $\chi_{\sigma}$= $0.18 ~t$. The doping level was
set such that (i) the self-consistently determined $S_z$ vanishes
without the impurity and (ii) the chemical potential is higher
than the vHS energy. The requirement of (i) ensures that the
moments do not exist $without$ the Zn impurity, and (ii) ensures
that the tunneling conductance background has a negative slope
around the Fermi level. In the present HFB mean-field
calculation, these two requirements are met around $40-55~\%$
doping for $t'=-0.45~t$ depending on $U$ and $J$. This extremely
high doping concentration, an artifact of the mean-field
calculation which could be cured by going beyond the mean-field
theory, was caused by the requirements of negative tunneling
conductance slope and the large (realistic) $t'$. These
constraints were not properly considered previously. Although the
present calculations can also be done around 15 \% doping without
the constraints, they were done at a high doping level with the
experimental constraints satisfied. The reason is that the
induced moments are not affected much by the largeness of the
doping level since they are mainly determined by the competition
of the electron kinetic energy and the effective exchange
integral. This has been checked by direct numerical calculations
without the requirements, and is further strengthened by the
satisfactory agreement of the computed NMR spectra from the
induced moments with experiments as is discussed in Fig.\ 3. On
the other hand, the resonant energy and peak strength depend on
the parameters of the model like the $t'$, $V_{imp}$, and doping
concentration somewhat sensitively as noted previously
\cite{martin}, and some care must be taken before comparing
different calculations.

We took $V_{imp} =
-100 ~t$ to model a Zn impurity placed at ${\vi }$ = (15,15) in a
$30 \times 30$ unit cell. Fig.\ 2 shows the staggered magnetic
moments $\tilde{S_z } (\vi ) = (-1)^{(i_x +i_y )} S_z (\vi )$
induced by the Zn. Antiferromagnetic moments develop
with the peaks at the four nearest sites around the impurity site.
The total spin $\sum_i S_z (\vi) = \frac{1}{2} \hbar$.
The magnetic moment vanishes at the impurity site, and the Zn
remains non-magnetic. The Friedel-like concentric oscillation with a period of
$3-4$ lattice constants can clearly be seen. The oscillation exhibits an
angular anisotropy as expected; along the anti-nodal direction
it gets more suppressed because of the pairing gap and along the nodal
direction it shows more pronounced oscillation.
The $d$-wave pairing gap, $\wt \Delta_d (\vi) = [\Delta_{\vi,\vi+x}
+\Delta_{\vi,\vi-x} -\Delta_{\vi,\vi+y} -\Delta_{\vi,\vi-y} ]/2$,
remains 0.312 $t$ for most of unit cell, but on the
impurity site $\wt \Delta_d (\vi)$ vanishes. The superconductivity
is severely destroyed by the non-magnetic impurity within about 2
lattice constants around the impurity site.

\begin{figure}
\includegraphics*[scale=0.8]{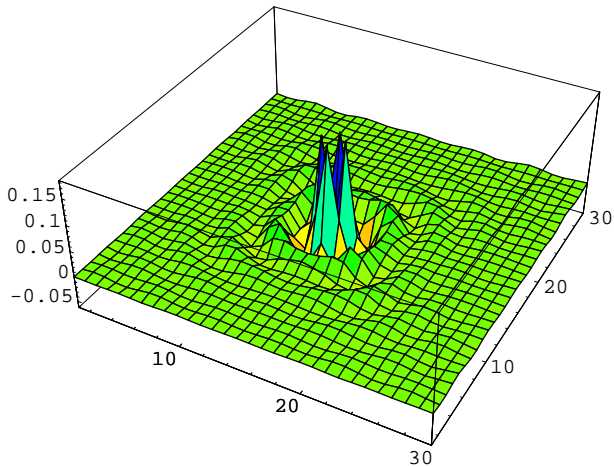}
\caption{ The staggered magnetic moments
$\tilde{S_z } (\vi )$ induced around an non-magnetic
impurity at $T=0.0005~t$, for $J=t$, $t'= -0.45~t$, and $U=3~t$.
Note that the induced moments reside
mainly on the 4 nearest neighbors and exhibit a Friedel-like oscillation.}
\end{figure}

From the Zn impurity-induced $S_z (\vi )$ shown in Fig.\ 2, we
have also calculated the broadened NMR spectra which may be
compared with the experimental $^{63}$Cu NMR spectra. The NMR
spectra is just a histogram of the frequency shifts given by $
h_i = A_{\alpha}^{onsite} S_z (\vi) + B \sum_{\tau} S_z (\vi
+\vec \tau ) $, where $A_{\alpha}^{onsite} = -13.8 ~T/\mu_B$ and
$B = 3.45 ~T/\mu_B$ for the magnetic field $\vec H \parallel
c$-axis \cite{julien}. The $T$-dependence of the induced moments
were calculated for several different $T$ with a $20\times 20$
unit cell. The distribution of the induced moments in $20\times20$
unit cell is almost the same with the $30\times 30$ one because
the magnetic correlation length is shorter than the unit cell
size. The results at $T=0.1~T_c$, $0.2~T_c$, $0.5~T_c$, and
$1.5~T_c$ are shown in Fig.\ 3. The width of the NMR spectra gets
narrower as $T$ is increased because the induced moments are
decreased. Compared with the NMR spectra [Fig.\ 1(a) in Ref.\
\cite{julien}], the results are quite satisfactory, which gives a
reliability to the present calculation of the induced magnetic
moments.

\begin{figure}
\includegraphics*[scale=0.33]{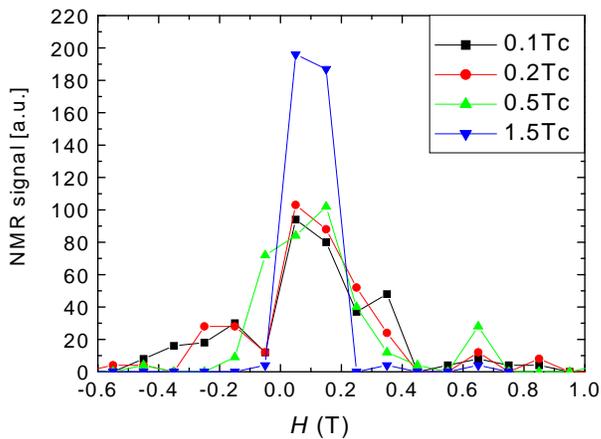}
\caption{ The Cu NMR spectra calculated from the distribution of
the induced magnetic moments due to a Zn impurity as shown in
Fig.\ 2. }
\end{figure}

There are two factors that contribute to the induction of the
magnetic moments by a non-magnetic impurity: the localization of
an electron and destruction of superconductivity around the
impurity site due to the deep impurity potential of the Zn. The
magnetic moment in a normal state is determined by the competition
between the effective exchange integral of $2J + U$ and the
electron hopping motion of $t$. When an electron is localized by
the impurity potential, the electron hopping gets suppressed,
which favors the appearance of magnetic moments around the
impurity site, as was studied by Odashima and Matsumoto in a
normal state \cite{odashima}. In a SC state, the appearance of
magnetic moments can also be suppressed by the onset of
superconductivity. Therefore, when the superconductivity is
destroyed by an impurity, magnetic moments can appear around the
impurity site. Both of above two factors contribute to the
appearance of the magnetic moments induced by the Zn impurity in
a SC state. Similar ideas were suggested by various groups in the
context of spin gaped systems \cite{kilian,sachdev2}. The idea
was that the freed spins created by an impurity abide by the
impurity site because of the confinement provided by
antiferromagnetic correlations and show up around the impurity.
The present theory can readily be applied to the magnetic
impurity effects such as the recent STM experiment on Ni
substituted \bscco \cite{hudson}. We are currently carrying out
such calculations and will report the results elsewhere.

To summarize, we considered the \ttju Hamiltonian to study the
non-magnetic Zn impurity effects in $d$-wave superconductivity.
The Bogoliubov-de Gennes equation in the real space was derived
from the Hamiltonian using the Hartree-Fock-Bogoliubov
approximation, and was solved self-consitently with the $d$-wave
pairing, bond, charge, and spin order parameters included. It was
demonstrated that a non-magnetic impurity induces the
antiferromagnetic moments around itself in a $d$-wave
superconductive state. The calculated NMR spectra due to the
induced moments are found to be in good agreement with the
experiments. It was also proposed that the negative slope of the
tunneling conductance may most probably come from the next-nearest
hopping integral $t'$.

We would like to thank Sasha Balatsky for helpful discussions.
This work was supported by the Korea Science \& Engineering
Foundation (KOSEF) through the grant No.\ 1999-2-11400-005-5, by
the Korea Research Foundation (KRF) through the grant No.\ KRF-2000-DP0139,
and by the Ministry of Education through the Brain Korea 21 SNU-SKKU Physics
Program.



\begin{thebibliography}{99}

\bibitem{ginsberg} D. M. Ginsberg, {\it Physical Properties 
of High Temperature Superconductors} Vol. I - V (World Scientific).
\bibitem{pan} S. H. Pan {\it et al.}, Nature (London) {\bf 403}, 746 (2000).
\bibitem{alloul} H. Alloul {\it et al.}, cond-mat/9905424.
\bibitem{julien} M. H. Julien {\it et al.}, Phys. Rev. Lett. {\bf
84}, 3422 (2000). 
\bibitem{bobroff} J. Bobroff {\it et al.}, Phys. Rev. Lett. {\bf 86}, 4116
(2001).
\bibitem{williams} G. V. M. Williams, J. L. Tallon, and R. Dupree,
Phys. Rev. B {\bf 61}, 4319 (2000).
\bibitem{balatsky} I. Martin, A. V. Balatsky, and J. Zaanen,
cond-mat/0012446.
\bibitem{ting2} J. X. Zhu, C. S. Ting, and C. R. Hu, Phys. Rev. B {\bf 62},
6027 (2000).
\bibitem{sachdev} A. Polkovnikov, S. Sachdev, and M. Vojta, Phys.
Rev. Lett. {\bf 86}, 296 (2001).
\bibitem{rantner} W. Rantner and X. G. Wen, Phys.
Rev. Lett. {\bf 85}, 3692 (2000).
\bibitem{balatsky2} A. V. Balatsky, M. I. Salkola, and A.
Rosengren, Phys. Rev. B {\bf 51}, 15547 (1995).
\bibitem{salkola} M. I. Salkola, A. V. Balatsky, and D. J. Scalapino,
Phys. Rev. Lett. {\bf 77}, 1841 (1996).
\bibitem{salkola2} M. I. Salkola, A. V. Balatsky, and J. R. Schrieffer,
Phys. Rev. B {\bf 55}, 12648 (1997).
\bibitem{chchoi} C. H. Choi, Phys. Rev. B {\bf 63}, 064507 (2001).
\bibitem{daul} S. Daul, D. J. Scalapino, and S. R. White,
Phys. Rev. Lett. {\bf 84}, 4188 (2000).
\bibitem{yaz} A. Yazdani {\it et al.} Phys. Rev. Lett. {\bf 83}, 176 (1999).
\bibitem{gimm} T. H. Gimm {\it et al.}, to be published in Physica C.
\bibitem{martin} I. Martin and A. V. Balatsky, cond-mat/0003142.
\bibitem{odashima} S. Odashima and H. Matsumoto, Phys. Rev. B {\bf
56}, 126 (1997).
\bibitem{kilian} R. Kilian, S. Krivenko, G. Khaliullin, and P.
Fulde, Phys. Rev. B {\bf 59}, 14432 (1999).
\bibitem{sachdev2} S. Sachdev and M. Vojta, cond-mat/0009202.
\bibitem{hudson} E. W. Hudson {\it et al.}, Nature (London) {\bf 411}, 920
(2001); M. E. Flatte, {\it ibid.}, 901 (2001).

\end{thebibliography}
\end{document}